\documentstyle[eqsecnum,aps,prd]{revtex}
\begin{document}
\newcommand{\bfx}{{\bf x}}
\newcommand{\bfy}{{\bf y}}
\newcommand{\bfr}{{\bf r}}
\newcommand{\bfk}{{\bf k}}
\newcommand{\bfa}{{\bf a}}
\newcommand{\bkp}{{\bf k'}}
\newcommand{\order}{{\cal O}}
\newcommand{\beq}{\begin{equation}}
\newcommand{\eeq}{\end{equation}}
\newcommand{\beqn}{\begin{eqnarray}}
\newcommand{\eeqn}{\end{eqnarray}}
\newcommand{\lmk}{\left(}
\newcommand{\rmk}{\right)}
\newcommand{\lkk}{\left[}
\newcommand{\rkk}{\right]}
\newcommand{\lnk}{\left\{}
\newcommand{\rnk}{\right\}}
\newcommand{\call}{{\cal L}}
\newcommand{\calh}{{\cal H}}
\newcommand{\ppp}{\partial}
\newcommand{\tilq}{{\tilde q}}
\newcommand{\tilp}{{\tilde p}}
\newcommand{\tilu}{{\tilde U}}
\newcommand{\tilh}{{\tilde H}}
\newcommand{\baret}{{\bar \eta}}
\newcommand{\xbar}{{\overline x}}
\thispagestyle{empty}
\thispagestyle{empty}
{\baselineskip0pt
\leftline{\large\baselineskip16pt\sl\vbox to0pt{\hbox{DAMTP} 
               \hbox{University of Cambridge}\vss}}
\rightline{\large\baselineskip16pt\rm\vbox to20pt{
\hbox{DAMTP/R/98/25}
               \hbox{UTAP-192}
               \hbox{\today}
\vss}}%
}
\vskip15mm
\begin{center}
{\large\bf Generation of Magnetic Field due to Excited Q-Balls}
\end{center}

\begin{center}
{\large Tetsuya Shiromizu} \\
\vskip 3mm
\sl{DAMTP, University of Cambridge \\ 
Silver Street, Cambridge CB3 9EW, UK \\
\vskip 5mm
Department of Physics, The University of Tokyo, Tokyo 113-0033, Japan \\
and \\
Research Centre for the Early Universe(RESCEU), \\ The University of Tokyo, 
Tokyo 113-0033, Japan
}
\end{center}

\begin{abstract} 
We investigate phase transitions due to excited Q-balls. 
As excited Q-balls have angular momentum, a magnetic field 
can be generated if one considers gauged Q-balls. 
Based on the course of the phase transition 
we estimate the strength of the magnetic field and then we 
find that it might be the origin of observed magnetic fields of 
astrophysical objects such as galaxies and clusters of galaxies. 
\end{abstract}
\vskip1cm


\section{Introduction}

The generation of the seed of the galactic magnetic fields for galaxies 
and clusters is
one of the most crucial problems in cosmology because one knows from 
magnetohydrodynamics that magnetic fields 
cannot exist if the initial field strength is zero. But now 
the coherent magnetic field over galaxies is observed as
 $\sim 10^{-6}{\rm Gauss}$\cite{Sofue}. 

If seed magnetic fields with 
$10^{-19}{\rm Gauss}$ exist over the present comoving scale of the proto 
galaxy, $\sim 100{\rm kpc}$, one expects that the galactic magnetic fields 
can be amplified by the dynamo mechanism\cite{Dynamo}. 

Recently, some strong candidates for the generation mechanism have 
been proposed in the electroweak phase transition. One of them is based 
on thermal fluctuations\cite{Thermal}. 
However, the coherent scale is too small 
and then the net magnetic field will be too small for the present observed 
magnetic fields. Another mechanism is to rely on bubble nucleation
\cite{Bubble}. In this case, the coherent length becomes large and then 
the field strength is enough for the onset of the dynamo mechanism. 
However, one needs to assume that the phase transition is strongly 
first order and completed by the bubble nucleation and expansion. Although 
several studies have been attempted, the detail of the phase
transition has not been clear at present\cite{Shiromizu}.

  
So we propose a new mechanism. We consider phase transitions for a 
complex scalar field. 
The concrete particle for 
the complex scalar field can be called a squark. 
As we consider general cases in this paper, 
we do not assign a concrete particle. 

As complex scalar fields has a conserved charge, one can expect that the 
phase transition differs from that of real scalar fields. In fact, 
there exist stable non-topological solitons: Q-balls\cite{Coleman}.
The generation of Q-balls\cite{Soliton} and the effect on phase 
transitions\cite{PT}\cite{Kusenko} have been actively 
investigated. However, these 
studies are focused on the ground state of the Q-ball. In the actual phase 
transition, excited Q-balls with surface waves might play an important role. 
Hence we shall reexamine the phase transition by taking account of the 
excitations. In such cases, the Q-ball has angular momentum and then gauged 
Q-balls\cite{Lee} with 
magnetic moment will be induced if, for example, the charge of the 
Q-ball is baryon number. Thus one can expect that 
a magnetic field will be generated in general. 

For simplicity, we consider the case of Q-balls which do not couple 
with gauge fields in this paper. 
Properly speaking, if one wants to consider the generation of the 
magnetic field, one should investigate using the gauged Q-ball. 
Thus, the present estimation of the magnetic field 
will be guessed from the non-gauged Q-balls, but it is sufficient 
for the order estimation. 

The rest of the present paper is organised as follows. In Sec. II, 
we construct the total energy of excited Q-balls.  In Sec. III, the 
phase transition will be considered. In Sec. IV, the typical strength 
of the generated magnetic field is estimated and then the net magnetic
 fields over the galaxy scale. 

\section{The total Energy of Excited Q-Balls}

In this section we examine the typical quantities of the excited Q-balls.
The total energy is given by 
%
\begin{equation}
E=\frac{Q^2}{2\int \varphi^2 d^3x}+\frac{1}{2}\int (\nabla \varphi)^2d^3x
+\int U(\varphi) d^3x,
\end{equation}
%
where we put the complex field $\psi$ as $\psi = e^{i \omega t}\varphi$ and 
$Q= \omega\int d^3x \varphi^2$. $\varphi$ is a time independent 
real function. Using the radius $R$, the wall width $\delta$, the field 
value at the asymmetric phase $\sigma_+$ and the field value at the top of the 
potential barrier $\sigma_-$, the terms on the right-hand side are 
given by 
%
\begin{equation}
\int \varphi^2 d^3x \simeq \frac{4\pi}{3}R^3\sigma_+^2, 
\end{equation}
%
%
\begin{equation}
\int (\nabla \varphi)^2 \simeq 4\pi R^2\delta\Bigl( \frac{\sigma_+}{\delta} 
\Bigr)^2[1+\ell (\ell +1)]
\end{equation}
%
and
%
\begin{equation}
\int U(\varphi) d^3x \simeq \frac{4\pi}{3}R^3U(\sigma_+)+4 \pi R^2\delta 
U(\sigma_-),
\end{equation}
%
respectively. Here we note that the contribution from the surface 
tension contains the perturbation 
of the mode $\ell$ in order to treat excited Q-balls.  
For brevity, we use the notation $U_\pm:=U(\sigma_\pm)$ hereafter. 
Substituting these expression into eq. (2.1), one can obtain 
%
\begin{equation}
E=\frac{3}{8\pi} \frac{Q^2}{R^3\sigma_+^2}+2\pi R^2\delta \Bigl( 
\frac{\sigma_+}{\delta}\Bigr)^2[1+\ell (\ell +1)]+4\pi R^2 \delta 
U_- +\frac{4\pi}{3}U_+R^3.
\end{equation}
%
To estimate the typical quantities one must seek the most probable configuration. 
The configuration will be decided by the variational principle. From 
$\partial E/\partial \delta|_{\delta=\delta_*}=0$, the wall width has the 
expression 
%
\begin{equation}
\delta_*=\frac{1}{{\sqrt {2}}} \frac{\sigma_+}{{\sqrt {U_-}}}[1+\ell(\ell+1)]
^{1/2}
\end{equation}
%
Thus, the total energy becomes
%
\begin{equation}
E=\frac{3}{8\pi}\frac{Q^2}{R^3\sigma_+^2}
+4 {\sqrt {2}}\pi [1+\ell(\ell+1)]^{1/2}\sigma_+{\sqrt {U_-}}R^2
+\frac{4\pi}{3}U_+R^3
\end{equation}
%
Let $T_c$ be the critical temperature at which the two 
vacua degenerate($U(0)=U_+=0$).
At $T>T_c$, as $ U_+>0$, one can see that the Q-ball is bounded inside finite 
radius and a stable solution exists. On the other hand, under 
the critical temperature, the Q-ball can evolve to a macroscopic 
size. In particular,  
it is worth noticing that there exists a critical charge $Q_c$. 
If the charge exceeds the critical one, Q-balls must expand for 
any energy. If the charge is smaller than the critical one, 
there exists a range of the total energy in which the 
Q-ball is bounded. In the next section, we examine the evolution of the phase transition 
as the temperature of the universe cools down.

\section{Phase Transition due to Excited Q-Balls}

First we consider the case of the critical temperature because  
one can expect that 
the difference between the phenomenon at the critical temperature and 
one at the higher temperature is qualitatively similar.  
At $T=T_c$, as $U_+$ vanishes, the total energy becomes
%
\begin{equation}
E=\frac{3}{8\pi}\frac{Q^2}{R^3\sigma_+^2}+4 {\sqrt {2}}\pi 
[1+\ell(\ell+1)]^{1/2}\sigma_+{\sqrt {U_-}}R^2
\end{equation}
%
>From $\partial E/ \partial R|_{R=R_*}=0$ one obtains the radius 
at which the energy is minimum,
%
\begin{equation}
R_*=\Bigl(\frac{9}{64{\sqrt {2}}\pi^2}\Bigr)^{1/5}[1+\ell(\ell+1)]^{-1/10}
\frac{Q^{2/5}}{\sigma_+^{3/5}U_-^{1/10}}.
\end{equation}
%
Thus the total energy becomes
%
\begin{equation}
E_*=E(R_*)=\frac{15}{16 \pi} \Bigl( \frac{64{\sqrt {2}}\pi^2 }{9} \Bigr)^{3/5}
[1+\ell(\ell+1)]^{3/10}\frac{Q^{4/5}U_-^{3/10}}{\sigma_+^{1/5}}.
\end{equation}
%
On the other hand, if the net charge is given by  
%
\begin{equation}
Q_* \simeq \Bigl(n_\varphi\frac{4\pi}{3}R_*^3\Bigr)^{1/2}
\end{equation}
%
one can obtain the expression 
%
\begin{equation}
Q_*=\Bigl( \frac{4\pi}{3}\Bigr)^{5/4} \Bigl( \frac{9}{64{\sqrt {2}}\pi^2} 
\Bigr)^{3/4} \eta_\varphi^{5/4}[1+\ell(\ell+1)]^{-3/8}
\frac{n_\gamma^{5/4}}{\sigma_+^{9/4}U_-^{3/8}},
\end{equation}
%
where $n_\varphi$ is the number density of $\psi$ particles 
and $\eta_\varphi=n_\varphi/n_\gamma$. 
Using this expression, the final expression of the total energy is given by
%
\begin{equation}
E_*=\frac{5}{4}\frac{n_\varphi}{\sigma_+^2}
\end{equation}
%
As $\beta E_* < 1$, the Q-ball is generated and then the 
probability will be proportional to $e^{-\beta E_*}$. 

Next, we consider the cases in which the temperature is lower than 
the critical one. As we stated in the previous section, a typical 
charge $Q_c$ exists. Hence, we concentrate on the critical case. 
For this case also, various quantities are decided by the variational principle:
%
\begin{equation}
\partial E/ \partial R|_{R=R_c,Q=Q_c}=\partial^2E/\partial R^2|_{R=R_c,Q=Q_c}= 0 .
\end{equation}
%
The results are as follows;
%
\begin{equation}
Q_c \simeq 34.0 [1+\ell(\ell+1)]^{3/2}\frac{\sigma_+^4}{U_-}
\Bigl( \frac{U_-}{|U_+|}\Bigr)^{5/2},
\end{equation}
%
%
\begin{equation}
R_c \simeq 2.36 [1+\ell(\ell+1)]^{1/2}\frac{\sigma_+}{ {\sqrt {U_-}} }
\frac{U_-}{|U_+|} 
\end{equation}
%
and
%
\begin{equation}
E_c \simeq 54.8 [1+\ell(\ell+1)]^{3/2}\sigma_+ \frac{\sigma_+^2}
{ {\sqrt {U_-}} } \Bigl( \frac{U_-}{|U_+|} \Bigr)^2.
\end{equation}
%

Following Coleman's estimation of surface waves the frequency\cite{Coleman} 
is written as 
%
\begin{equation}
(k^0_\ell)^2=\frac{\alpha}{\rho_0 R_0^3}\ell(\ell+2)(\ell-1)
\end{equation}
%
where $\alpha =\int^{\sigma_+}_0d \varphi [2U-\omega^2\varphi^2]^{1/2}$. 
As $\alpha \simeq {\sqrt {2U_-}}\sigma_+$ and $\rho_0 R_c^3 
\simeq \frac{3}{4\pi}E_c^{\ell=0}$, the frequency becomes 
%
\begin{equation}
(k^0_\ell)^2 \simeq 0.11\frac{U_-}{\sigma_+^2}
\Bigl( \frac{|U_+|}{U_-} \Bigr)^2\ell(\ell+2)(\ell-1)
\end{equation}
%
and then the rotation velocity $v_\ell $ is given by 
%
\begin{equation}
v_\ell \simeq R_c^{\ell=0}k^0_\ell \simeq 0.78 [\ell(\ell+2)(\ell-1)]^{1/2}.
\end{equation}
%
Assuming conservation of angular momentum, one 
can derive the final rotation velocity as follows,
%
\begin{eqnarray}
v_f & = & \Bigl(\frac{R_f}{R_c}\Bigr)^{1/2}v_\ell \nonumber \\
    & \simeq & 0.66 f_b^{1/2}g_*^{-1/4}{\sqrt {\ell(\ell+2)(\ell-1)}} 
\Bigl( \frac{m_{\rm pl}}{T_f}\Bigr)^{1/2}\Bigl( \frac{|U_+|}{U_-} \Bigr)^{1/2}
\Bigl( \frac{U_-^{1/2}}{\sigma_+ T_f}  \Bigr)^{1/2},
\end{eqnarray}
%
where $R_f$ is the final radius of the Q-ball as the phase transition 
finishes and given by $ \sim f_b H^{-1}$. 

Even if  $Q_* < Q_c$ at $T>T_c$, the Q-ball will attain critical charge soon 
by charge accretion at the temperature $T<T_s<T_c$\cite{Kusenko}. 
Hence, when the temperature cools down such that $(T_c -T)/T_c \ll 1$, 
the Q-ball can expand. But the fraction of such Q-balls might be too small. 
The probability $P(\ell)$ of a Q-ball with the surface wave of mode $\ell$ is 
given by 
%
\begin{eqnarray}
P(\ell) \sim e^{-\beta E_c}
\end{eqnarray}
%
because the probability is decided by $P(\ell) \sim e^{-\beta E_*}$ at 
the critical temperature. 

The phase transition proceeds by the expanding Q-ball with probability 
$P(\ell)$\footnote{In the analysis using gauged Q-balls, one 
must consider the contribution from the gauge field. However, the 
order of the magnitude is the same as that from the $\ell=2$ surface wave 
of the scalar field. Thus, one can expect that the probability 
$P(\ell)$ does not have drastic changes from that of the 
gauged Q-ball. The detailed discussion will be appear 
in Ref. \cite{izumi}}. A Q-ball with surface waves has 
angular momentum. Thus, 
if one considers a gauged Q-ball, the excited gauged Q-ball has non-zero 
magnetic moment and a magnetic field can be generated. In the next section 
we estimate the order of the magnitude of the magnetic field.

\section{Generation of Magnetic Fields}

Now, we can estimate the magnetic field. As we stated in Sec. I, 
excited Q-balls have angular momentum and then have a magnetic 
field if one considers gauged Q-balls. 
Thus one can observe the generation of a magnetic field. As 
one must consider the gauged Q-ball, note that the Q-ball has an upper 
bound on the total charge. The maximum value $Q_{\rm max}$ is given by 
%
\begin{eqnarray}
Q_{\rm max} =\frac{1}{{\sqrt {2}}}\frac{\pi}{e^4}
\end{eqnarray}
%
for $e < 2^{-3/2}$\cite{Lee}. In this paper, as we stated in the 
introduction, we estimate the typical order 
of the generated magnetic field only using the non-gauged excited Q-ball 
in the previous section, because the above value $Q_{\rm max}$ 
is not greatly different from the critical charge 
$Q_c$ when $e \sim 2^{-3/2}$ and $\ell=2$. Hence, this extrapolation 
cannot destroy the present basic idea on the generation mechanism.

The magnetic moment is given by 
%
\begin{eqnarray}
M_\ell & \simeq & \frac{Q_c}{R_f^3}v_f \times R_f \times R_f^3 \nonumber \\
       & = & Q_c v_f R_f
\end{eqnarray}
%
Thus the magnetic field becomes
%
\begin{equation}
B_\ell \sim \frac{M_\ell}{R_f^3}\sim 7.76 f_b^{-3/2}g_*^{3/4}[1+\ell(\ell+1)]^{3/2}
[\ell(\ell+2)(\ell-1)]^{1/2} T_f^2  
\Bigl( \frac{U_-}{|U_+|}\Bigr)^2 \Bigl( \frac{T_f}{m_{\rm pl}}\Bigr)^{3/2}
\Bigl( \frac{\sigma_+}{U_-^{1/4}}\Bigr)^3 \Bigl( \frac{\sigma_+}{T_f}\Bigr)^{1/2}
\end{equation}
%
Averaging over the ensemble one can obtain the order 
%
\begin{eqnarray}
B \sim {\sqrt {\langle B^2 \rangle}} & = & 
{\sqrt { \frac{\sum_\ell B^2_\ell {\rm exp}[-\beta E_c] }
{\sum_\ell {\rm exp}[-\beta E_c]} }} \nonumber \\
& \simeq & 4.1 \times 10^2 f_b^{-3/2}g_*^{3/4}T_f^2  
\Bigl( \frac{U_-}{|U_+|}\Bigr)^2 \Bigl( \frac{T_f}{m_{\rm pl}}\Bigr)^{3/2}
\Bigl( \frac{\sigma_+}{U_-^{1/4}}\Bigr)^3 \Bigl( \frac{\sigma_+}{T_f}\Bigr)^{1/2}
{\rm exp}\Bigl[-\frac{\beta}{2}(E_c^{\ell=2}-E_c^{\ell=0})\Bigr] \nonumber \\
& \sim & 4.1 \times 10^2 f_b^{-3/2}g_*^{3/4} T_f^2 
\Bigl( \frac{T_f}{\sigma_+}\Bigr)^{1/2}\Bigl( \frac{\sigma_+}{U_-^{1/4}}\Bigr)
\Bigl( \frac{T_f}{m_{\rm pl}}\Bigr)^{3/2} \gamma
e^{-480 \gamma},
\end{eqnarray}
%
where we defined
%
\begin{eqnarray}
\frac{\beta}{2}[E_c^{\ell=2}-E_c^{\ell=0}]& \simeq & 480 \beta \sigma_+
\frac{\sigma_+^2}{{\sqrt {U_-}}}\Bigl( \frac{U_-}{|U_+|}\Bigr)^2 \nonumber \\
& =: & 480\gamma
\end{eqnarray}
%
The above mean value is maximum at $\gamma_* \sim 2.1 \times 10^{-3}$ 
and the maximum value is $B_{\rm max} \sim 4.5 \times 10^2 
(T_f/100{\rm GeV})^{7/2}{\rm Gauss}$ for 
$f_b \sim 10^{-3}$, $g_* \sim 100$ and $T_f \sim U^{1/4}_-\sim \sigma_+$. 

As the Reynolds number will becomes 
${\rm Re} \sim 10^{12}$\cite{Bubble} at the 
endpoint of the phase transition, turbulence for the electroweak plasma 
occurs over the scale of the domain radius and then the energy of the 
magnetic field is equipartitioned with the 
energy of the fluid\footnote{Although the amplification due to the
turbulence is not clear at present, it has been used in the 
generation mechanism of the primordial magnetic field during 
the first order phase transition\cite{Bubble}. }. 
Hence the final field strength becomes
%
\begin{eqnarray}
B(R_f)\sim \rho^{1/2} v \sim g_*^{1/2} v T_f^2 \sim 10^{24}
\Bigl( \frac{T_f}{100{\rm GeV}}\Bigr)^2{\rm Gauss},
\end{eqnarray}
%
where $\rho$ and $v$ are the density and the velocity of the fluid, 
respectively. 

If the coherent scale is comoving, the number of magnetic domains inside 
the galaxy scale is $N \sim 10^{10}(10^{-3}/f_b)$. Thus, averaging over the 
galaxy scale\cite{EO}, one can obtain the present mean value 
%
\begin{eqnarray}
\langle B_{\rm now} \rangle_{\rm galaxy} & \sim & 
\frac{1}{N^{3/2}}\Bigl(\frac{a_f}{a_{\rm now}}  \Bigr)^2 B({R_f}) 
\nonumber \\
& \sim & 10^{-21}\Bigl( \frac{f_b}{10^{-3}}\Bigr)^{3/2} 
\Bigl( \frac{T_f}{100{\rm GeV}}\Bigr)^{-3}{\rm Gauss}.
\end{eqnarray}
%
The above value satisfies the onset condition of the dynamo mechanism. 

Finally we give some concluding remarks. First, we estimated the 
magnetic field using non-gauged Q-ball. However, properly speaking, 
one should do it using a gauged Q-ball. But, before performing a rough 
analysis using 
the gauged excited Q-ball, one should consider the fundamentals of the 
non-gauged and gauged excited Q-ball. 
The perturbation analysis will be reported in the next study\cite{izumi}. 
In the analysis, we find that the model $\ell=1$ perturbation exists. 

Second, we did not study 
deeply the dynamics of the phase transition due to Q-balls. 
One needs the information of 
the final stage in the phase transition. The analysis will be done 
by numerical simulations. 

Third, we did not take account of 
the details of the evolution such as diffusion and reconnection 
of the magnetic field after recombination, and 
damping\cite{Jedamik} due to the  
heat conductivity and shear viscosity before recombination. 
Although there does not exist a complete work 
for such kinds of physical processes, some progress exists. For example, 
Olsen et al showed that an inverse cascade happens if the power of the 
initial spectral is larger than $-3$\cite{Pattern} and this means that 
the coherent scale of the magnetic field extends.

\vskip1cm

\centerline{\bf Acknowledgment}
The author thanks Katsuhiko Sato and Yasushi Suto for their encouragements. 
He also would like to thank Sean A. Hayward for his careful reading of
this manuscript and Masahide Yamaguchi for his discussion. Finally he 
is grateful to Gary Gibbons and DAMTP relativity group for their
hospitality.  
The work is partially supported by a JSPS fellowship. 



\begin{thebibliography}{22}
\bibitem{Sofue}
Y. Sofue, M. Fujimoto and R. Wielebinski, Ann. Rev. Astron. Astrophs. {\bf 24}
(1986), 459
\bibitem{Dynamo}
For example, E. N. Parker, Cosmological Magnetic Field (Oxford Univ. 
Press, Oxford, 1979)
\bibitem{Thermal}
T. Vachaspati, Phys. Lett. {\bf B265}(1991),258.
\bibitem{Bubble}
G. Baym, D. Bodeker and L. McLerran, Phys. Rev. {\bf D53}(1996),662
\bibitem{Shiromizu}
For example, 
T. Uesugi, T. Shiromizu and M. Morikawa, Prog. Theor. Phys. {\bf 96}, 377
(1996);\\
K. Kajantie, M. Laine, K. Rummukainen and 
M. Shaposhnikov,  Nucl.Phys. {\bf B493},413(1997) and reference therein 
\bibitem{Coleman}
S. Coleman, Nucl. Phys. {\bf B262}(1985),263
\bibitem{Lee}
K. Lee, J. A. Stein-Schabes, R. Watkins and L. M. Widrow, Phys. Rev. 
{\bf D39}, 1665(1989)
\bibitem{Soliton}
K. Griest and E. W. Kolb, Phys. Rev. {\bf 40}(1989),3231;\\
J. A. Frieman, A. V. Olinto, M. Gleiser and C. Alcock, Phys. Rev. {\bf D40}, 
(1989),3241;\\
K. Griest, E. W. Kolb and A. Massarotti, Phys. Rev. {\bf 40}(1989),3529
\bibitem{PT}
D. Spector, Phys. Lett. {\bf B194}(1987),103;\\
J. Ellis, K. Enqvist, D. V. Nanopoulos and K. A. Olive, Phys. Lett. 
{\bf B225}(1989),313
\bibitem{Kusenko}
A. Kusenko, Phys. Lett. {\bf B406}(1997),26 
\bibitem{EO}
K. Enqvist and P. Olsen, Phys. Lett. {\bf B319}(1993),178
\bibitem{Jedamik}
K. Jedamzik, V. Katalinic and A. V. Olinto, Phys. Rev. {\bf D57} (1998), 3264
\bibitem{Pattern} 
P. Olsen, Phys. Lett. {\bf B398}(1997),321;\\
A. Brandenburg, K. Enqvist and P. Olsen, Phys. Lett. {\bf B391}(1997),395
\bibitem{izumi}
T. Shiromizu, T. Uesugi and M. Aoki, in preparation
\end{thebibliography}
\end{document}